\newcommand\showchange{TT}
\pdfoutput=1
\documentclass[format=acmsmall,10pt]{acmart}
\makeatletter 
\renewcommand\@formatdoi[1]{\ignorespaces} 
\makeatother 
\setcopyright{none}

\usepackage{import}
\usepackage{amsthm}
\usepackage{threeparttable}
\usepackage{amsmath}
\usepackage{amssymb}
\usepackage{multirow}
\usepackage{enumitem}
\usepackage{xspace}
\usepackage{listings}
\usepackage[misc]{ifsym}
\usepackage{fancyhdr}
\usepackage{pgfplots}
\pgfplotsset{compat=1.12}
\usepackage{subfigure}
\usepackage{filecontents}
\usepackage{caption}
\usepackage[english]{babel}
\usepackage{blindtext}
\usepackage{graphicx}
\usepackage{etoolbox}
\usepackage{natbib}
\newcommand{\cell}[1]{\begin{tabular}[c]{@{}l@{}}#1\end{tabular}}

\usepackage[ruled,linesnumbered]{algorithm2e} 

\SetAlFnt{\small}
\SetAlCapFnt{\small}
\SetAlCapNameFnt{\small}
\SetAlCapHSkip{0pt}
\IncMargin{-\parindent}

\usepackage{soul}	
\setstcolor{red}

\long\def\com#1{}

\if\showchange

\else

\fi

\def\bibbrev#1#2{#1}			

\newcommand{\bibconf}[3][]{#1 \bibbrev{#2}{#3 (#2)}}

\newcommand{\blind}[1]{}		

\newcommand{\mysect}[1]{\paragraph{#1}}		
\newcommand{\kw}[1]{{\em#1}}	
\newcommand{\kn}[1]{\texttt{#1}}	

\newcommand\etal{\emph{et al.\ }}
\newcommand\eg{\emph{e.g.,\ }}
\newcommand\ie{\emph{i.e.,\ }}

\theoremstyle{definition}

\newcommand\fullname{{\bf CO}ntext-sensitive and {\bf D}uration-{\bf A}ware 
{\bf R}emapping algorithm~({\sc Codar})\xspace}
\newcommand\name{{\sc Codar} remapper\xspace}
\newcommand\mysys{{\sc Codar}\xspace}
\newcommand\qamfname{{\bf M}ulti-architecture {\bf A}daptive {\bf Q}uantum {\bf A}bstract {\bf M}achine 
 ({\sf maQAM})\xspace}
\newcommand\qam{{\sf maQAM}\xspace}

\usepackage[braket]{qcircuit}
\usepackage{tikz}
\usetikzlibrary{calc}
\usetikzlibrary{backgrounds}
\usetikzlibrary{math}

\newcommand{\Nat}[0]{$\mathbb{N}$\xspace}
\newcommand{\Intmax}[0]{\textsf{INT\_MAX}\xspace}
\newcommand{\Arrow}[2]{{#1 {$\!\rightarrow\!$} #2}}

\newcommand{\NPQubit}[0]{N\xspace}
\newcommand{\NPGate}[0]{M\xspace}
\newcommand{\NLQubit}[0]{n\xspace}
\newcommand{\NLGate}[0]{m\xspace}
\newcommand{\NSeqs}[1]{\{1,2,..., #1\}\xspace}
\newcommand{\QArch}[0]{$\mathbb{A}$\xspace}
\newcommand{\QArchS}[0]{$\mathsf{A_s}$\xspace}
\newcommand{\QArchD}[0]{$\mathsf{A_d}$\xspace}
\newcommand{\LQubit}[0]{$\mathsf{Q_P}$\xspace}          
\newcommand{\PQubit}[0]{$\mathsf{Q_H}$\xspace}          
\newcommand{\PGate}[1]{$\mathsf{G}_{{#1}}$\xspace}      
\newcommand{\lqubit}[1]{{$q_{{#1}}$\xspace}}            
\newcommand{\pqubit}[1]{{$Q_{{#1}}$\xspace}}            
\newcommand{\lqubitvars}[0]{{\lqubit{\NSeqs{\NLQubit}}\xspace}}    
\newcommand{\pqubitvars}[0]{{\pqubit{\NSeqs{\NPQubit}}\xspace}}    
\newcommand{\pgate}[1]{{\textsf{g}$_{{#1}}$\xspace}}    
\newcommand{\lgate}[1]{{$g_{{#1}}$\xspace}}             
\newcommand{\pgatevars}[0]{{\pgate{\NSeqs{\NPGate}}\xspace}}    
\newcommand{\lgatevars}[0]{{\lgate{\NSeqs{\NLGate}}\xspace}}    
\newcommand{\lgates}[1]{{$I_{{#1}}$\xspace}}            
\newcommand{\lgatesschd}[1]{{$\mathcal{E}_{{#1}}$\xspace}}      
\newcommand{\gname}[1]{{\kn{gate(#1)}\xspace}}          
\newcommand{\gqubits}[1]{{\kn{qseq(#1)}\xspace}}        
\newcommand{\QCoupling}[0]{$\mathbb{M}$\xspace}
\newcommand{\QDistance}[0]{$\mathbb{D}$\xspace}
\newcommand{\PQEdge}[0]{$\mathsf{E_H}$\xspace}          
\newcommand{\QCycle}[0]{$\tau_u$}
\newcommand{\GDuration}[1]{{$\tau_{\textsf{#1}}$\xspace}}   

\newcommand{\qmap}[1]{{$\pi_{#1}$\xspace}}

\newcommand{\qlock}{$t_{end}$\xspace}

\usepackage[normalem]{ulem}
\begin{document}
\title{%
Context-Sensitive and Duration-Aware 
Qubit Mapping for Various NISQ Devices
}

\author{Yu Zhang~\Letter}
\thanks{This work was partly supported by the grants of the National Natural
Science Foundation of China (No. 61772487),
Anhui Initiative in
Quantum Information Technologies (No. AHY150100) and Anhui Provincial Major
Teaching and Research Project (No. 2017jyxm0005).\\
Corresponding author: Yu Zhang, email: yuzhang@ustc.edu.cn.
\\
{\em This is the extended abstract of the talk accepted by the First International Workshop on Programming Languages for Quantum Computing (PLanQC 2020), Jan 19, 2020, New Orleans, Louisiana, USA
}}
\email{yuzhang@ustc.edu.cn}
\author{Haowei Deng}
\email{jackdhw@mail.ustc.edu.cn}
\author{Quanxi Li}
\email{crazylqx@mail.ustc.edu.cn}

\affiliation{
\institution{University of Science and Technology of China}
\city{Hefei}
\country{China}
}

\begin{abstract}
Quantum computing (QC) technologies have reached a second renaissance in the last decade.
Some fully programmable QC devices have been built based on superconducting or ion trap technologies.
Although different quantum technologies have their own parameter indicators, 
QC devices in the NISQ era share common features and challenges such as 
limited qubits and connectivity,
short coherence time and 
high gate error rates.
Quantum programs written by programmers could hardly run on real hardware directly since two-qubit gates are usually allowed on few pairs of qubits.
Therefore, quantum computing compilers must resolve the mapping problem and transform original programs to fit the hardware limitation.\par

To address the issues mentioned above,
we summarize different quantum technologies
and abstractly define Quantum Abstract Machine (QAM);
then propose a \fullname based on the QAM.
By introducing lock for each qubit, \mysys is aware of gate duration difference and program context, which bring it abilities to extract more program's parallelism and reduce program execution time.
Compared to the best-known algorithm, 
\mysys halves the total execution time of several quantum algorithms and 
cut down 17.5\% $\sim$ 19.4\% total execution time on average in different architectures.
\end{abstract}

\maketitle

      \thispagestyle{fancy} 
      \lhead{} 
      \chead{} 
      \rhead{} 
      \lfoot{} 
      \cfoot{\thepage} 
      \rfoot{} 
      \renewcommand{\headrulewidth}{0pt} 
      \renewcommand{\footrulewidth}{0pt} 
\pagestyle{fancy}
      \cfoot{\thepage}
\section{Introduction}
\label{sec:intro}

Quantum Computing (QC) has attracted huge attention in recent a decade 
due to its ability to exponentially accelerate several important algorithms~\cite{nielsen2010quantum:qcqi, shor1997primefact,grover1996search,deutsch1992rapid}. 
Both QC algorithm designers and programmers work at a very high level,
and need to know little about (future) Noisy Intermediate-Scale Quantum (NISQ) devices 
that (will) execute quantum programs.
There exists a gap, however, 
between  NISQ devices and the hardware %
requirements (\eg size and reliability) of QC algorithms. 
To bridge this gap, 
QC requires abstraction layers and 
toolchains to translate and optimize applications~\cite{chong2017pl}.
QC compilers typically translate high-level QC code into 
(optimized) circuit-level assembly code in multiple stages.
In order to use NISQ hardware,
quantum circuit programs have to be compiled to the target device,
which includes mapping logical qubits to physical ones of the device.
The mapping step, 
which we focus on in this abstract,
faces a tough challenge 
because further physical constraints 
have to be considered.
In fact,
2-qubit gates can only be applied to certain physical qubit pairs.
Therefore,
additional SWAP operations have to be inserted 
in order to ``move''
the logical qubits to positions where they can interact with each other.
This qubit mapping problem has been proved to be a NP-Complete problem~\cite{siraichi2018qubitalloc}.

Previous solutions to this problem 
can be classified into two types.
One type is to formulate the problem into 
an equivalent mathematical problem and apply a solver
~\cite{maslov2008quantum,chakrabarti2011linear,shafaei2013optimization,shafaei2014qubit,wille2014optimal,lye2015determining,venturelli2017temporal,venturelli2018compiling,booth2018comparing,oddi2018greedy,bhattacharjee2017depth,wille2019mapping},
and another type is to use heuristic search to obtain approximate results
\cite{saeedi2011synthesis,lin2014paqcs,wille2016look,shrivastwa2015fast,kole2016heuristic,kole2018new,bhattacharjee2018novel,zulehner2019mapping,li2019tackling}.
The former suffers from very long runtime 
and can only be applied to small size cases.
The latter is better in runtime especially when the circuit is in a large scale. 
All of them assume 
different gates have the same execution duration.\par

\begin{table*}[!h]
\centering
\scriptsize
\caption{Parameter information of several quantum computing devices.}  
\label{tab:qc-args}
\begin{tabular}{@{}c@{}|@{}c|@{}c|@{}c|@{}c|@{}c|c|c@{}}
\hline
\multicolumn{2}{c|}{} &  
\multicolumn{2}{c|}{Ion Trap} &
\multicolumn{3}{c|}{Superconducting} & \multirow{2}{*}{Neutral Atom\cite{sheng2018PRL121.240501}}  
\\
\cline{3-7}
\multicolumn{2}{c|}{}
& Ion Q5{\cite{linke2017QCcomparison}}
& Ion Q11{\cite{wright2019benchmarking}}
& IBM Q5{\cite{linke2017QCcomparison}} 
& IBM Q16{\cite{murali2019noise}}
& IBM Q20{\cite{li2019tackling}}
\\
\hline
\multicolumn{2}{c|}{Available 1-qubit gate}
& \multicolumn{2}{c|}{$R^\theta_\alpha$} &\multicolumn{3}{c|}{X, Y, Z, H, S, T} & $R^\theta_\alpha$
\\
\multicolumn{2}{c|}{Available 2-qubit gate}
& \multicolumn{2}{c|}{XX} &\multicolumn{3}{c|}{CNOT} &CNOT{\cite{stockill2017phase}}
\\
\hline
\multirow{4}{*}{Fidelity} 
& 1-qubit gate & 99.1(5)\% &99.5\%& 99.7\%  & $\sim$99.8\% & $\sim$99.56\% &99.995\% {\cite{sheng2018PRL121.240501}}
\\ 
& 2-qubit gate & 97(1)\% &97.5\%[95.1\%,98.9\%]& 96.5\% & $\sim$96\% & $\sim$97\% 
& 82\%\cite{maller2015rydberg}
\\ 
& 1-qubit readout & \ket{0}:99.7(1)\%, \ket{1}:99.1(1)\% &99.3\%& $\sim$ 96\% & $\sim$93\% & $\sim$91.2\% & 98.6\% \cite{fuhrmanek2011free}
\\ 
& average readout & 95.7(1)\% &--& $\sim$ 80\% & -- &--  &  $>$97\% 
\cite{martinez2017fast}\\
\hline

\multirow{2}{*}{Time} & 1-qubit gate & 20$\mu$ s&&130 ns&80 ns& --& 1$\mu$ $\sim$20$\mu$ s
\\
&2-qubit gate & 250$\mu$ s&--& 250-450 ns &170-391 ns& --
& $\sim$10$\mu$ s
\\
\hline

\multicolumn{2}{c|}{Depolarization ($T_1$)}
& $\sim \infty$  &--& $\sim$ 60$\mu$ s  & $\sim$ 70$\mu$ s & 87.29$\mu$ s  & $>$10s
\\
\multicolumn{2}{c|}{Spin dephasing ($T_2$)}
& $\sim$ 0.5s  &--& $\sim$ 60$\mu$ s & $\sim$ 70$\mu$ s& 54.43$\mu$ s & $\sim$ 1s  
\\
\hline
\end{tabular}
\end{table*}

On NISQ hardware,
however,
different gates have different durations (see Table~\ref{tab:qc-args}).
Ignoring the gate duration difference may cause
these algorithms to find the shortest depth but not the shortest execution time.
The real execution time of the circuit is associated with 
the weighted depth, 
in which different gates have different duration weights.
Considering gate duration difference will help the compiler 
make better use of the parallelism of quantum circuit
and generate the circuit with shorter execution time.
In this abstract,
we focus on solving the \kw{qubit mapping problem} by heuristic search with the consideration of gate duration difference and program context to explore more program's parallelism.
To address the challenges of qubit mapping problem
and adapt to different quantum technologies,
we first give several examples to explain our motivation,
then propose a quantum abstract machine (QAM) 
for studying the qubit mapping problem.
The QAM is modelled as a 2D coupling graph with limited connectivity
and configurable durations of different kinds of quantum gates.
Based on the QAM,
we further propose two mechanisms that enable \fullname to solve the qubit mapping problem with the awareness of gate duration difference and program context. Experimental results show that compared to the best
known remapping algorithm, \mysys can cut down 17.5\% $\sim$ 19.4\% weighted depth at average.

\section{Problem Analysis}
\label{sec:problem}

\subsection{Recent Work on Qubit Mapping}
\label{sec:problem:related}
There are a lot of research on the qubit mapping problem.
Here we focus on analyzing some valuable solutions in recent two years~\cite{siraichi2018qubitalloc,zulehner2019mapping,li2019tackling,wille2019mapping,murali2019noise,ash2019qure,tannu2019not}.
All of them 
are proposed for 
some IBM QX architectures,
and none of them consider the gate duration difference.

\paragraph{Solutions only considering qubit coupling}
\cite{siraichi2018qubitalloc,wille2019mapping} provide solutions for 5-qubit IBM QX architectures with directed coupling.
Siraichi \etal~\cite{siraichi2018qubitalloc} propose an optimal algorithm 
based on dynamic programming, 
which only fits for small circuits;
then they propose a heuristic one which is fast but oversimplified with results worse than IBM's solution.
Wille \etal~\cite{wille2019mapping} present a solution with a minimal number of additional \kn{SWAP} and \kn{H} operations,
in which qubit mapping problem is formulated as a symbolic optimization problem with high complexity. 
They utilize powerful reasoning engines to solve the computationally task.

\cite{zulehner2019mapping,li2019tackling} use heuristic search to provide good solutions 
in acceptable time for large scale circuits.
Zulehner \etal divide the two-qubit gates into independent layers, 
then use $A^*$ search plus heuristic cost function to determine compliant mappings for each layer~\cite{zulehner2019mapping}.
Li \etal propose a SWAP-based bidirectional heuristic search algorithm, named SABRE~\cite{li2019tackling}, 
which can produce comparable results with exponential speedup
against previous solutions such as~\cite{zulehner2019mapping}. 

\paragraph{Solutions further considering error rates}
\cite{ash2019qure,murali2019noise,tannu2019not} provide another type of perspective for solving the qubit mapping problem. 
They consider the variation in the error rates of different qubits and connections to generate directly executable circuits that improve reliability rather than minimize circuit depth and number of gates. 
Based on the error rate data from real 
IBM Q16 and Q20 respectively,
\cite{murali2019noise,tannu2019not} 
use a SMT solver to schedule gate operations to qubits with lower error probabilities. 
Ash-Saki \etal propose two approaches, Sub-graph Search and Greedy approach, to 
optimize gate-errors~\cite{ash2019qure}.
Circuits generated by them may suffer from long execution time due to no consideration of the minimal circuit depth.

\paragraph{What we consider in the qubit mapping}
We want to produce solutions for the qubit mapping problem with speedup against previous works and maintain the fidelity meanwhile. 
Besides the coupling map, 
what we further concern includes the program context and the gate duration difference,
which affect the design of qubit mapping.
Considering these factors will help to find remapping solution with approximate optimal execution close to reality.

\subsection{Motivating Examples}
\label{sec:problem:ex}
\begin{figure}[h]
\begin{minipage}[b]{.18\textwidth}
\centering
{
\includegraphics[width=\textwidth]{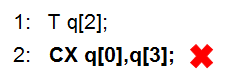}
\\~~
\includegraphics[width=0.4\textwidth]{grid4.png}
\\
{(a)}
}
\end{minipage}
\begin{minipage}[b]{.15\textwidth}
\centering
{\scriptsize
\begin{tabular}{c|c}
\hline
    Gate  & Duration
    \\
    \hline
    T & 1 cycle \\
    \hline
    CX & 2 cycle \\
    \hline
    SWAP & 6 cycle \\
    \hline
\end{tabular}
}
\\
{(b)}

\end{minipage}
\begin{minipage}[b]{.32\textwidth}
\centering
\includegraphics[width=\textwidth]{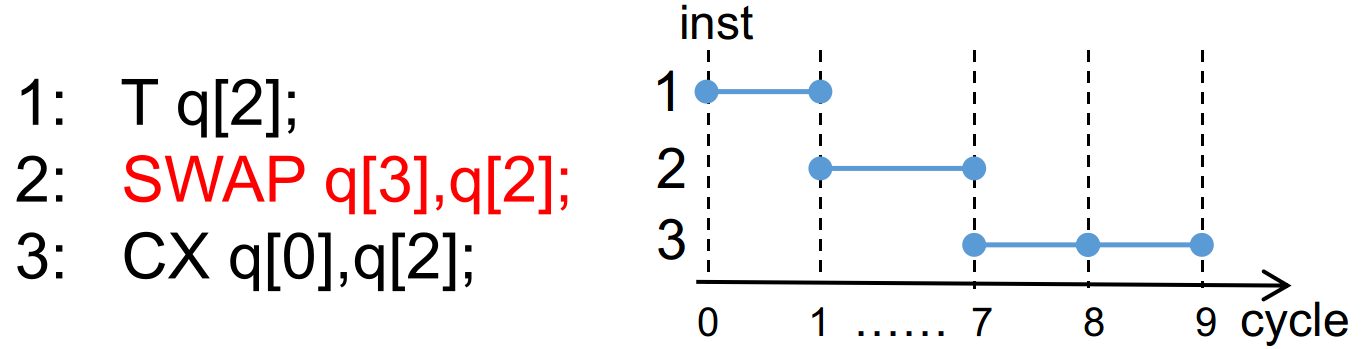}
\\
{(c)}
\end{minipage}
\begin{minipage}[b]{.32\textwidth}
\centering
\includegraphics[width=\textwidth]{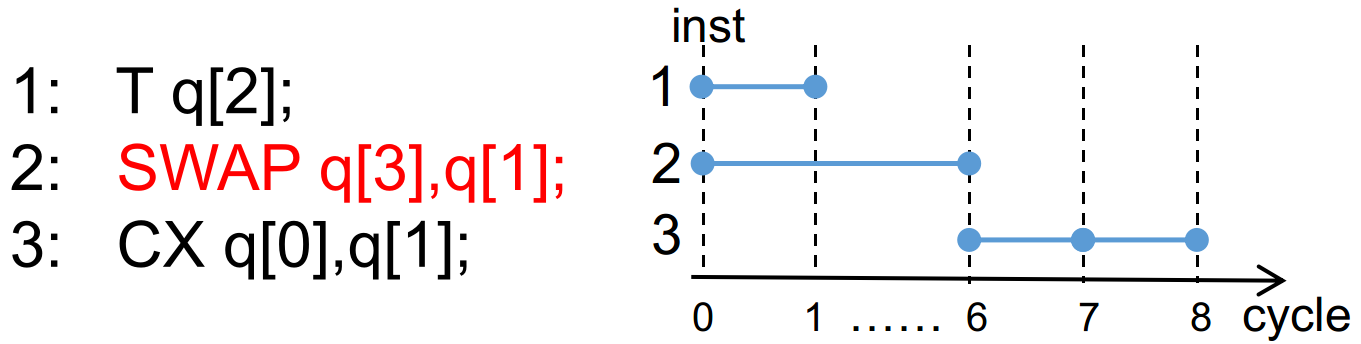}
\\
{(d)}
\end{minipage}
\caption{An example reflecting the impact of {\bf program context} on SWAP-based transformations: 
\kn{SWAP q[3], q[1]} is selected in (d) to avoid using \kn{q[2]} operated by the previous \kn{T} gate, accordingly increasing parallelism.}
\label{fig:ex1:qft4}
\end{figure}

We use several examples written in   
OpenQASM~\cite{cross2017OpenQASM}   
to explain our motivation for considering program context
and gate duration difference in qubit remapping process.
The two examples base on the coupling map of four physical qubits $Q_0\sim Q_3$ and 
the assumed gate durations 
defined in Fig.~\ref{fig:ex1:qft4} (a) and (b).
We directly map the logical qubits \kn{q[0]}$\sim$\kn{q[3]} initially 
to physical qubits $Q_0\sim Q_3$ for easier explanation.

\mysect{Impact of program context.}
\label{sec:problem:ex:context}

Consider the OpenQASM code fragment 
shown in Fig.\ref{fig:ex1:qft4} (a),
where \kn{CX} means \kn{CNOT} in OpenQASM.
Since qubits $Q_0$ and $Q_3$ are non-adjacent,
the instruction ``\kn{CX q[0],q[3]}'' at line 2 cannot be applied. 
To solve the problem,
\kn{SWAP} operation is required before performing the \kn{CX} operation.
In this case,
there are four candidate \kn{SWAP} pairs,
\ie ($Q_0$, $Q_1$), ($Q_0$, $Q_2$), ($Q_3$, $Q_1$) and ($Q_3$, $Q_2$). 
If the program context, \ie the predecessor instruction 
``\kn{T q[2];}'', is not considered, 
there are no differences among four candidates when selecting.
However, \kn{SWAP} operation on pair ($Q_3$, $Q_2$) or ($Q_0$, $Q_2$) conflicts with the context instruction ``\kn{T q[2];}'' due to operating the same $Q_2$,
and has to be executed serially after \kn{T} operation as shown in Fig.\ref{fig:ex1:qft4} (c).
\kn{SWAP} on pair ($Q_3$, $Q_1$) or ($Q_3$, $Q_1$) does not conflict with ``\kn{T q[2];}'' 
and can be executed in parallel as shown in Fig.\ref{fig:ex1:qft4} (d).
With the awareness of the context information, 
\kn{SWAP} operations which improve parallelism can be sifted out.
\begin{figure}[h]
\begin{minipage}[b]{0.07\textwidth}
\centering
\includegraphics[width=\textwidth]{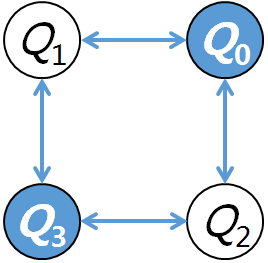}
\\
{(a)}
\end{minipage}
\begin{minipage}[b]{.17\textwidth}
\centering
\includegraphics[width=0.8\textwidth]{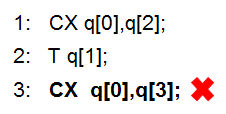}
\\
{(b)}
\end{minipage}
\begin{minipage}[b]{.36\textwidth}
\centering
\includegraphics[width=0.9\textwidth]{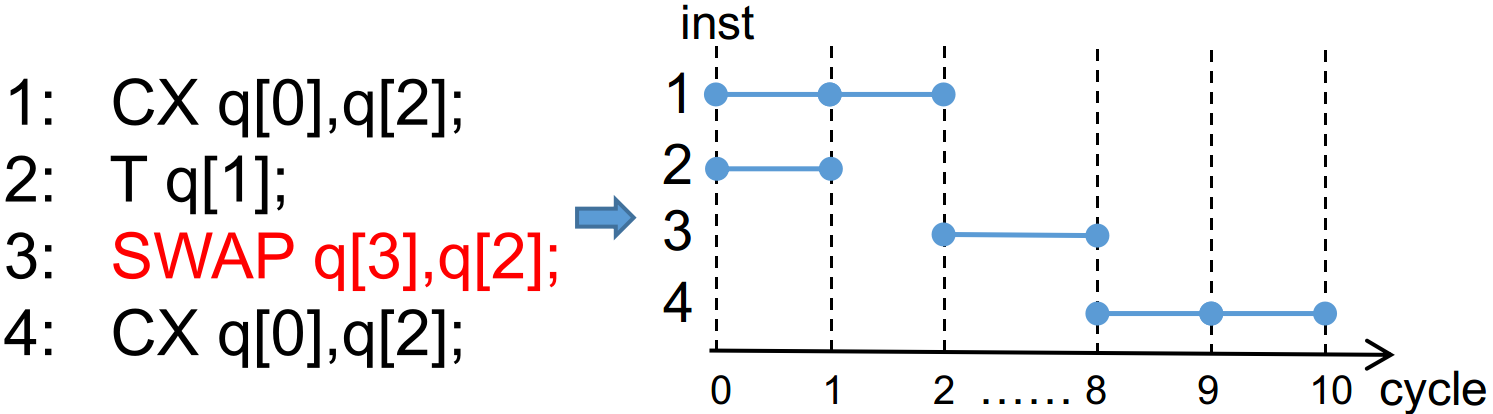}
\\~~
{(c)}
\end{minipage}
\begin{minipage}[b]{.36\textwidth}
\centering
 \includegraphics[width=0.9\textwidth]{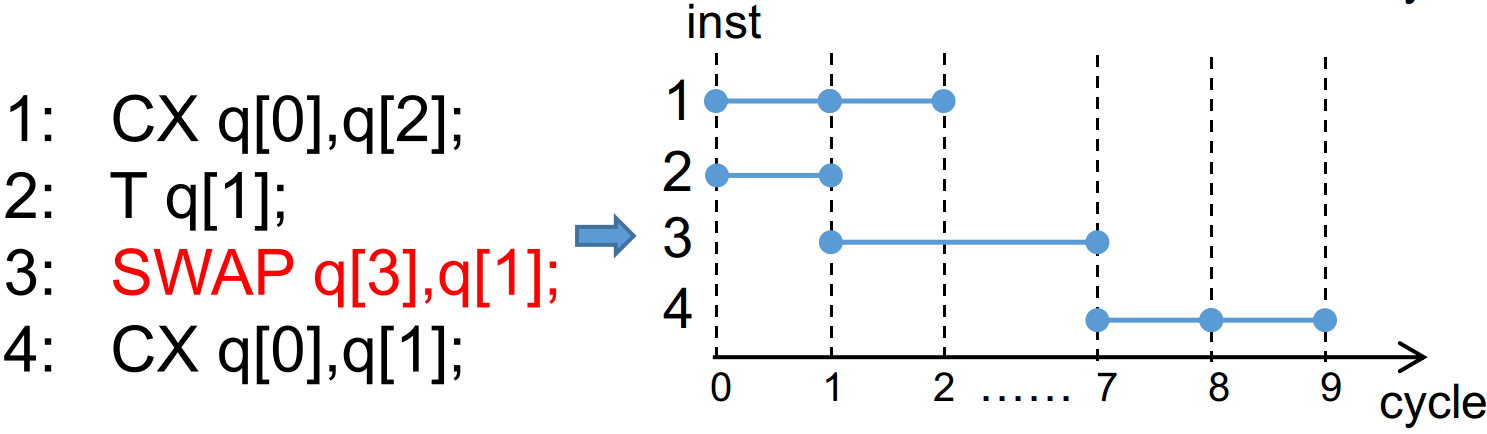}
\\{(d)}
\end{minipage}
\caption{A 4-qubit QFT example reflecting the impact of {\bf gate duration difference}: 
``\kn{SWAP q[3],q[1]}'' 
is the best candidate
since it can start immediately after
``\kn{T q[1]}'' while ``\kn{CX q[0],q[2]}'' has not finished yet, increasing the parallelism of the circuit.}
\label{fig:exdu:qft4}
\end{figure}

\mysect{Impact of gate duration difference.}
\label{sec:problem:ex:duration}
We use a 4-qubit QFT (quantum fourier transform) circuit to explain the limitation of ignoring the duration of quantum gates.
Fig.~\ref{fig:exdu:qft4} (b) lists the fragment of a 4-qubit QFT OpenQASM program,
which is generated by ScaffCC compiler~\cite{abhari2014scaffcc}.
Similar to the first example, 
\kn{SWAP} operation is required before performing the \kn{CX} operation and there are also the same four candidate \kn{SWAP} pairs. 
Instructions ``\kn{T q[2]}'' and ``\kn{CX q[0],q[2]}'' can be executed in parallel 
and we assume both of them start at cycle 0. 
If the difference of gate durations is ignored, 
the two gates ``\kn{T q[2]}'' and ``\kn{CX q[0],q[2]}'' are assumed to finish at the same time $t$ 
and the four candidate \kn{SWAP} operations have to start after $t$. 
But if the duration of \kn{CX} is twice as much as that of \kn{T},
we find that ``\kn{T q[2]}'' will finish at cycle 1 while ``\kn{CX q[0],q[2]}'' at cycle 2. 
As a result, \kn{SWAP} between \kn{q[3]} and \kn{q[1]} can start at cycle 1 as shown in Fig.\ref{fig:exdu:qft4} (d),
while other three candidate SWAP operations 
have to start at cycle 2
since one of operands $Q_0$ or $Q_2$ is occupied
as shown in Fig.\ref{fig:exdu:qft4} (c). 
Fig.~\ref{fig:exdu:qft4} (d) has better parallelism,
which can be deduced by the awareness of 
different quantum gate durations. 

\subsection{Quantum Architecture Abstraction}
\label{sec:model}
\begin{table*}[!h]
\centering
\scriptsize
\caption{Definition of Quantum Abstract Machine.}
\label{tab:qam:def}
\begin{tabular}{l|l|l} 
\hline
&
\textbf{Notation} & \textbf{Definition} \\

\hline 
\multirow{5}{*}{\cell{Static\\ Structure}}&
\PQubit & The set of physical qubits, $|$\PQubit$|$ = $\NPQubit$;
$\forall$\pqubit{}$\in$\PQubit, \pqubit{}.\qlock is the qubit lock described in Section~\ref{sec:design:lock}\\
\cline{2-3}
&   \PGate{} & The set of elementary quantum operations and \kn{SWAP}, $|$\PGate{}$|$ = $\NPGate$\\
\cline{2-3}
&   \QCoupling=(\PQubit,\PQEdge) & The coupling 
graph of a quantum device \\
\cline{2-3}
&   \GDuration{}: \Arrow{\PGate{}}{\Nat} & Mapping from quantum operations to their durations, 
\Nat represents the set of natural numbers\\
\cline{2-3}
&   \QDistance: \Arrow{\PQubit$\times$\PQubit}{\Nat} & Mapping from physical qubit pairs 
to their shortest path lengths on the \QCoupling{}, \\
&& if there is no path between \pqubit{i} and \pqubit{j}, then \QDistance(\pqubit{i},\pqubit{j}) = \Intmax  \\
\hline

\multirow{2}{*}{\cell{Dynamic \\Structure}}
&   \qmap{}:\Arrow{\LQubit}{\PQubit} & Mapping from logical qubits to physical qubits \\
\cline{2-3}

&   CF(\lgates{}) & Commutative Front gate set of a gate sequence \lgates{}, defined in Definition~\ref{def:cfgates} \\
\hline

\multirow{2}{*}{\cell{Auxiliary\\Functions}}
&   \gname{\lgate{}} &  the name of a given gate \lgate{}. \\
\cline{2-3}
&   \gqubits{\lgate{}} & the logical qubit sequence applied by a given gate \lgate{}. \\
\hline

\multirow{5}{*}{\cell{Variables}}
&   \pqubitvars & Physical qubits, \pqubit{i}$\in$\PQubit, $1\le i\le \NPQubit$\\
\cline{2-3}
&   \lqubitvars & Logical qubits,  \lqubit{i}$\in$\LQubit, $1\le i\le \NLQubit$ \\
\cline{2-3}
&   \pgatevars & Physical quantum operations, \pgate{i}$\in$\PGate, $1\le i\le \NPGate$\\
\cline{2-3}
&   \lgatevars & Quantum operations in the circuit program\\
\cline{2-3}
&   \lgates{}  & A sequence of quantum operations, \lgates{} = [\lgate{1},\lgate{2},..., \lgate{k}] if $k=|$\lgates{}$|$,
and the length of \lgates{} is written as \lgates{}$.len$\\
\hline
\end{tabular}
\end{table*}

Since the qubit mapping problem is 
affected by the constraints of underlying QC devices,
which base on various and evolving quantum technologies, 
it is essential to design quantum mapping algorithms that are compatible with different quantum technologies.

In view of the above, 
we consider the qubit connectivity of various NISQ devices, 
and take each gate duration as a multiple of quantum clock cycle \QCycle,
which can be analogized to the classic clock cycle.
We then introduce a \qamfname 
which consists of static and dynamic structures, 
denoted as \QArch = (\QArchS, \QArchD).
Table~\ref{tab:qam:def} shows the definitions for \qam,
where \QArchS = (\PQubit, \PGate{}, \QCoupling, \GDuration{}, \QDistance),
and \QArchD = (\qmap{}, CF).
We assume the device can provide enough physical qubits (denote the number as $\NPQubit$) for the program's execution 
(denote the number of logical qubits in the program as $\NLQubit$), \ie $\NPQubit\ge\NLQubit$.

For a QC device, we abstract its qubit layout as a 
graph \QCoupling
where qubits are vertices and 
there are 
edges between 
qubit pairs where a two-qubit gate is allowed to apply on them. 
We introduce the Gate Duration Map~\GDuration{} into \QArchS
which maps each kind of quantum gate to its duration, 
depending on the information from quantum architecture. 
We assume the same kind of quantum gates have the same duration and fidelity. 
We also introduce the shortest distance matrix map \QDistance between each pair of physical qubits 
for quick selection of exchangeable qubits in our \mysys scheduling algorithm.

\section{Design}
\label{sec:design}
In this section, 
we discuss our 
\fullname.
We first overview the idea of \mysys, 
then introduce the two key mechanisms that enable \mysys
context-sensitivity and duration awareness.

The main idea of \mysys is to generate an executable gate sequence for a given input OpenQASM program 
by adjusting the gate sequence and inserting the swap operation with the program semantics unchanged.
The generated gate sequence fits quantum hardware limitation on one hand,
and has better parallelism on the other hand to reduce the circuit's weighted depth , \ie simulated execution time.
We propose qubit lock mechanism in Section~\ref{sec:design:lock} for quickly finding available qubits.
And we adjust the gate order based on the quantum gate commutativity described in Section~\ref{sec:design:detect}.

\subsection{Qubit Lock}
\label{sec:design:lock}
\mysys is based on a reasonable assumption: 
\kw{a qubit cannot be applied by two or more gates at the same time}. 
If a qubit is occupied by a gate, it is called \kw{busy} (not free) qubit and cannot be applied by other gates. 
As the example shown in Fig.~\ref{fig:ex1:qft4}, 
when inserting \kn{SWAP} for a specific two-qubit gate \kn{CX q[0],q[3]}, 
the neighbour qubit \kn{q[2]} of the target qubits may be occupied by 
the contextual gate which has started in earlier time. 
Using the occupied qubits to route the two-qubit gate will reduce the parallelism of the program 
because the routing process has to wait until occupied qubits become free. 

\begin{figure}[htbp]
\centering
\begin{minipage}[b]{.35\textwidth}
    \centering
    \includegraphics[width=\textwidth]{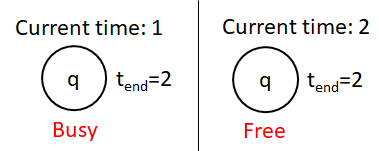}
    \caption{Qubit lock. A 2-cycle gate is applied on qubit q at time 0, 
    then q is busy until time 2. }
    \label{fig:lock}
\end{minipage}
\quad
\begin{minipage}[b]{.6\textwidth}
\centering
\includegraphics[width=\textwidth]{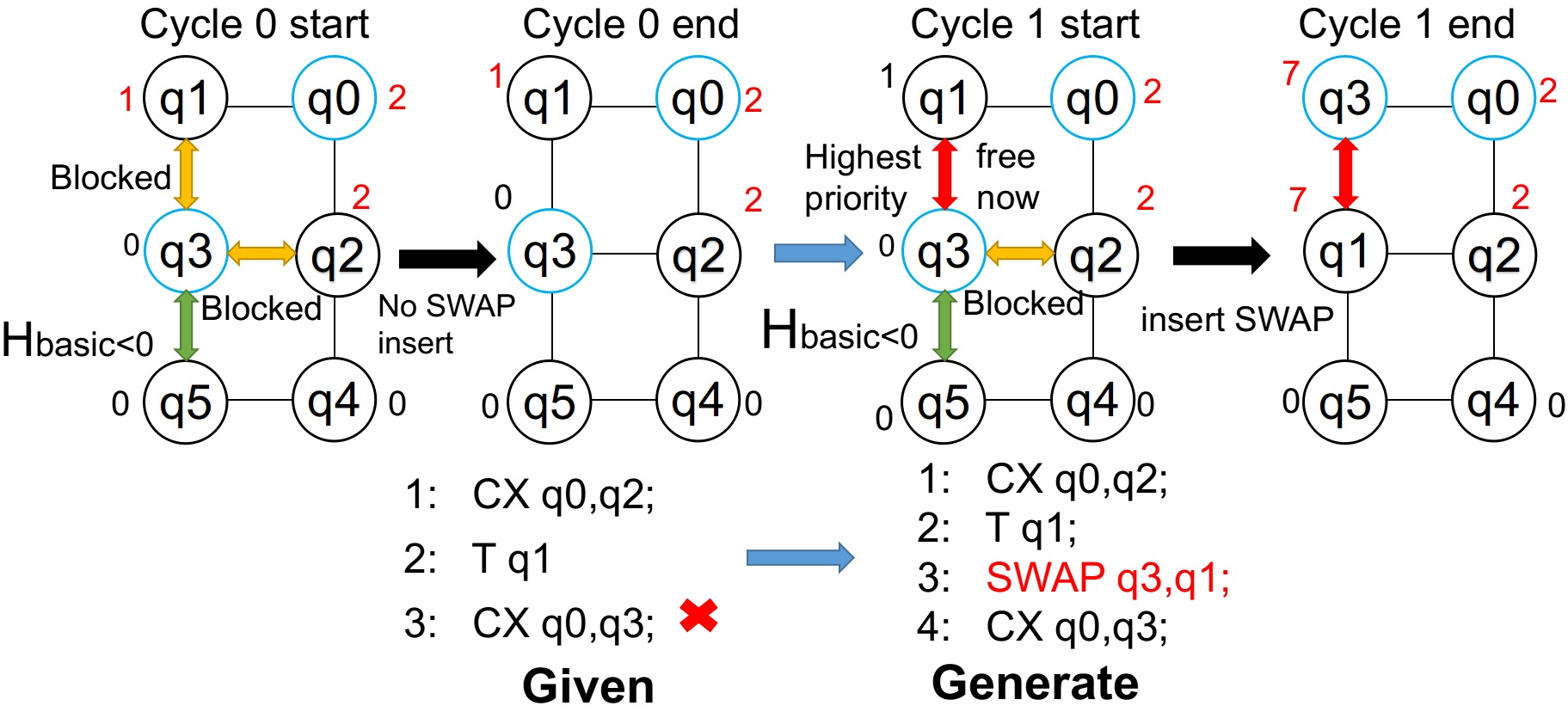}
\caption{Example of heuristic search for the high parallelism \kn{SWAP}. The number near the qubits denote the qubit lock \qlock.}
\label{fig:ex-alg} 
\end{minipage}
\quad
\end{figure}

To make \mysys aware of the qubit occupation by the past contextual gate, 
we introduce a {\bf qubit lock} \qlock for each physical qubit in \pqubit. 
When start applying a quantum gate \pgate{} $\in$ \PGate{} at time $t$ on a physical qubit in \pqubit{}  
and the gate's duration is \GDuration{\pgate{}}, 
\mysys will update this qubit's \qlock as $t$+ \GDuration{\pgate{}} which means 
that it is occupied before $t$+\GDuration{\pgate{}}. 
A qubit is \kw{free} only when its lock \qlock $\le$ current time. 
When try to find routing path for a specific two-qubit gate, 
by comparing \qlock of each qubit with the current time, 
\mysys can be aware of which qubit is occupied by the past contextual gate. 
Fig.~\ref{fig:lock} shows an example. 
Gates can only be applied to the physical qubits in free state. 
The gates whose associate physical qubits are all free, are called \kw{lock free} gates.

Qubit lock can also help \mysys aware of the gate duration difference. 
Different gate kinds have different durations 
and \mysys updates the operated qubit's lock \qlock with different value. 
As a result, 
qubits applied by gates with shorter duration will be set smaller \qlock 
and become free earlier. 
Thus \mysys can use those earlier free qubits to route two-qubit gates 
and improve the parallelism of the program. 
As the example shown in Fig.~\ref{fig:exdu:qft4}~(d), 
suppose the program starts at time 0 and \GDuration{T}=1, \GDuration{CNOT}=2. 
Then \qlock of \pqubit{1} is set to 1 while \qlock of \pqubit{0} and \pqubit{2} are set to 2. 
\pqubit{1} becomes free at time 1 while \pqubit{2} is still busy. 
\mysys can use \pqubit{1} to route for the third gate and need not wait for the freedom of \pqubit{2}.

\subsection{Commutativity Detection}
\label{sec:design:detect}
Qubit lock brings \mysys awareness of the past contextual gate. 
Considering gate commutation relation can expose more future contextual gate 
for \mysys to decide routing path.
\begin{definition}[Commutative Forward Gate, CF gate]
\label{def:cfgates}
    Given a gate sequence \lgates{}=[\lgate{1}, \lgate{2}, ..., \lgate{k}, ...], $\forall$ \lgate{k} $\in$ \lgates{},
    \lgate{k} is a commutative forward gate iff 
    $\forall j, 0 < j < k$, \lgate{j} and \lgate{k} are commutative. 
\end{definition}
    The commutation relation between two-qubit gates \lgate{A}, \lgate{B} 
    that share qubits with each other can be resolved 
    by checking the relevant unitary operators $\hat{A}\hat{B} = \hat{B}\hat{A}$. 
    Gates applied to disjoint qubits are obviously commutative with each other.
    If a commutative forward gate is commutative with all the gates before it in sequence \lgates{}, 
    it can exchange with the head of \lgates{}.
    
    All CF gates in sequence \lgates{} are denoted as $CF(I)$,
    which can be \kw{executed instantly} from software perspective.
    Compared to the method that 
    fetches gates with no predecessor as instantly-executable gates, 
    choosing CF gates as instantly-executable gates can expose more contextual gates for heuristic search 
    to determine better remapping solutions.
    
    Suppose a sequence \lgates{} contains two gates: \kn{CX q1,q3} and \kn{CX q2,q3} in order. 
    The second gate shares \kn{q3} with the first and 
    might not be regarded as instantly executable due to qubit dependence. 
    However, because the second commutes with the first 
    and is a CF gate in \lgates{}, 
    it is instantly executable in fact. 
    Commutativity detection will expose both \kn{CX}s for heuristic search 
    which will improve its contextual look-ahead ability. \par
\subsection{Example}
\label{sec:design:ex}

Now we use an example shown in Fig.~\ref{fig:ex-alg} to explain our algorithm. 
Suppose there is a 6-qubit device and we are given a gate sequence \lgates{} 
that contains a \kn{CX} on \{q0,q2\}, a \kn{T} on \{q1\} and a \kn{CX} on \{q0,q3\}. 
The number near the qubit node represents the value of its \qlock. 
All the three gates are CF gates. 
Due to the coupling limitation, \kn{CX} on \{q0,q3\} is not directly executable and 
\kn{SWAP} is needed. 
The algorithm simulates the execution timeline and starts at cycle 0. 
At cycle 0, the first gate "\kn{CX} q0,q2" and the second gate "\kn{T} q1" are directly executable 
so both of them will be launched and qubits \{q0,q1,q2\}'s \qlock locks are updated with the gate duration
(\kn{T}=1 cycle, \kn{CX}=2 cycle). 
At cycle 0, each of \{q0,q1,q2\} has bigger \qlock than current time 
and thus they are locked. 
Therefore the \kn{SWAP} between \{q1,q3\} and \{q2,q3\} are blocked. 
\kn{SWAP} between \{q3,q5\} with $H_{basic}<0$
(which means the \kn{SWAP} won’t shorten the total distance of CF gates according to our heuristic cost function)
moves q3 away from q0 and 
will not be inserted. 
As a result, no \kn{SWAP} will be inserted in cycle 0 and 
the mapping \qmap{} stays unchanged. 
At cycle 1, qubit q1 becomes free while q2 stays busy. 
The \kn{SWAP} between \{q1,q3\} becomes free while the \kn{SWAP} between \{q3,q2\} is still blocked. 
Therefore the algorithm will know that the \kn{SWAP} between \{q1,q3\} can start earlier than \kn{SWAP} between \{q3,q2\} 
and choose \kn{SWAP} q3,q1 to solve the remapping problem. 
After launching the \kn{SWAP} between \{q1,q3\}, 
qubit locks of \{q1,q3\} are also updated by the sum of its start time (cycle 1) and 
the duration of \kn{SWAP} (6 cycle) as 7.

\section{Experimental Evaluation}
\label{sec:eval}
In this section, 
we evaluate \mysys with benchmarks based on the latest reported hardware models. 
\mysect{Comparison with Previous Algorithms}
Several recent algorithms proposed by IBM~\cite{qiskit},
Siraichi $\etal$\cite{siraichi2018qubitalloc}, 
Zulehner $\etal$\cite{zulehner2019mapping} and Li $\etal$\cite{li2019tackling} try to find solutions of the qubit mapping problem with small circuit depth. 
Among them, 
Li's SABRE~\cite{li2019tackling} beats the other three in the performance of benchmarks,
thus it is used for comparison in this paper. \mysect{Hardware Configuration}
We test our algorithm on several latest reported architectures, including IBM Q20 Tokyo\cite{li2019tackling}, IBM Q16 Melbourne\cite{ibmqdeviceinfo}, $6\times6$ grid model
proposed by Enfield~\cite{siraichi2018qubitalloc}'s GitHub and Google Q54 Sycamore~\cite{arute2019google54}. 
The gate duration difference configuration is based on experimental data of symmetric superconducting technology 
shown in Table~\ref{tab:qc-args}, 
where two-qubit gate duration is generally twice as much as that of the single-qubit gate. 
\mysect{Benchmarks}
To evaluate our algorithm, we totally collect 71 benchmarks which are selected from the previous work, including:
1) programs from IBM Qiskit~\cite{cross2018ibmqiskit}'s Github and  RevLib~\cite{wille2008revlib};
2) several quantum algorithms compiled from ScaffCC~\cite{abhari2014scaffcc} and Quipper~\cite{green2013quipper};
3) benchmarks used in the best-known algorithm SABRE~\cite{li2019tackling}. 
The size of the benchmarks ranges from using 3 qubits up to using 36 qubits and about 30,000 gates.
For the IBM Q16, Q20 and $6\times6$ architectures,
68 benchmarks out of the 71 benchmarks except 3 36-qubit programs are tested. 
While all 71 benchmarks are tested on Google Q54 Sycamore. 
%
%

\newcommand{\figwidth}{4.4cm}
\newcommand{\figheight}{3.4cm}
\begin{figure}[h]
\centering
\scriptsize
\begin{minipage}[b]{0.23\textwidth}
\centering
\begin{tikzpicture}
\begin{axis}[
ybar,
title=IBM Q16 Melbourne,
width=\figwidth,
height=\figheight,
enlargelimits=0.03,
xticklabels=\empty,
ybar interval=0.2,
ymin=0.5,
]
\draw[red,thick] (axis cs:0,1) -- (axis cs:71,1);
\addplot coordinates {
(1,1.279)(2,1.214)(3,1.490)(4,1.221)(5,1.465)
(6,1.175)(7,1.000)(8,1.200)(9,1.457)(10,1.482)
(11,1.283)(12,1.138)(13,1.262)(14,1.164)(15,1.053)
(16,1.813)(17,1.411)(18,1.833)(19,1.371)(20,1.109)
(21,1.731)(22,1.298)(23,2.038)(24,1.379)(25,1.144)
(26,1.312)(27,2.103)(28,1.304)(29,1.011)(30,1.120)
(31,1.108)(32,2.031)(33,1.205)(34,1.099)(35,1.099)
(36,1.103)(37,1.921)(38,1.316)(39,1.101)(40,0.989)
(41,1.256)(42,1.241)(43,1.167)(44,1.106)(45,1.138)
(46,1.096)(47,1.110)(48,1.925)(49,1.113)(50,1.109)
(51,1.000)(52,1.101)(53,1.568)(54,1.039)(55,2.021)
(56,1.232)(57,1.098)(58,1.091)(59,1.312)(60,1.941)
(61,1.105)(62,1.253)(63,1.306)(64,1.380)(65,1.050)
(66,1.403)(67,1.062)(68,1.000)
};

\end{axis}
\end{tikzpicture}
\end{minipage}
~
\begin{minipage}[b]{0.23\textwidth}
\centering
\begin{tikzpicture}
\begin{axis}[
ybar,
title=Enfield 6$\times$6,
width=\figwidth,
height=\figheight,
enlargelimits=0.03,
xticklabels=\empty,
ybar interval=0.2,
ymin=0.5,
]
\draw[red,thick] (axis cs:0,1) -- (axis cs:71,1);
\addplot coordinates {
(1,1.028)(2,1.214)(3,1.174)(4,1.059)(5,1.134)
(6,1.180)(7,1.001)(8,0.947)(9,1.456)(10,0.912)
(11,1.174)(12,1.069)(13,1.045)(14,1.029)(15,0.964)
(16,1.437)(17,1.229)(18,1.835)(19,1.234)(20,1.946)
(21,0.932)(22,2.151)(23,0.915)(24,1.222)(25,1.059)
(26,1.578)(27,1.072)(28,1.012)(29,1.087)(30,1.079)
(31,1.089)(32,1.822)(33,1.359)(34,1.087)(35,2.033)
(36,1.336)(37,1.154)(38,1.217)(39,1.247)(40,1.263)
(41,0.895)(42,0.976)(43,1.017)(44,1.096)(45,1.091)
(46,1.839)(47,1.255)(48,1.110)(49,1.271)(50,0.967)
(51,1.098)(52,1.866)(53,1.184)(54,2.025)(55,1.386)
(56,1.136)(57,1.308)(58,1.123)(59,2.605)(60,1.413)
(61,1.004)(62,1.213)(63,1.354)(64,1.194)(65,1.059)
(66,1.252)(67,0.922)(68,1.102)
};
\end{axis}
\end{tikzpicture}
\end{minipage}
~
\begin{minipage}[b]{0.23\textwidth}
\centering
\begin{tikzpicture}
\begin{axis}[
ybar,
title=IBM Q20 Tokyo,
width=\figwidth,
height=\figheight,
enlargelimits=0.03,
xticklabels=\empty,
ybar interval=0.2,
ymin=0.5,
]
\draw[red,thick] (axis cs:0,1) -- (axis cs:71,1);
\addplot coordinates {
(1,0.673)(2,0.786)(3,1.298)(4,1.312)(5,1.036)
(6,0.975)(7,1.000)(8,1.200)(9,1.348)(10,1.000)
(11,0.989)(12,0.711)(13,1.000)(14,0.796)(15,0.937)
(16,1.214)(17,1.540)(18,1.400)(19,1.337)(20,1.057)
(21,1.353)(22,1.509)(23,1.632)(24,1.417)(25,1.389)
(26,1.202)(27,1.476)(28,1.497)(29,1.011)(30,1.011)
(31,0.997)(32,1.435)(33,1.633)(34,1.019)(35,1.016)
(36,1.026)(37,1.640)(38,1.588)(39,1.380)(40,1.488)
(41,1.232)(42,1.475)(43,1.122)(44,1.435)(45,1.312)
(46,1.054)(47,1.051)(48,1.519)(49,1.640)(50,1.019)
(51,1.000)(52,1.062)(53,1.690)(54,1.397)(55,1.735)
(56,1.636)(57,0.964)(58,1.090)(59,1.045)(60,1.639)
(61,1.765)(62,1.503)(63,1.715)(64,1.520)(65,1.083)
(66,1.792)(67,1.512)(68,1.978)
};
\end{axis}
\end{tikzpicture}
\end{minipage}
\quad
~
\begin{minipage}[b]{0.23\textwidth}
\centering
\begin{tikzpicture}
\begin{axis}[
ybar,
title=Google Q54 Sycamore,
width=\figwidth,
height=\figheight,
enlargelimits=0.03,
xticklabels=\empty,
ybar interval=0.2,
ymin=0.5,
]
\draw[red,thick] (axis cs:0,1) -- (axis cs:71,1);
\addplot coordinates {
(1,1.139)(2,1.063)(3,1.404)(4,1.051)(5,1.451)
(6,1.202)(7,1.000)(8,1.125)(9,0.885)(10,1.107)
(11,1.174)(12,1.114)(13,1.152)(14,0.897)(15,1.053)
(16,1.188)(17,1.411)(18,1.688)(19,1.404)(20,1.091)
(21,1.115)(22,1.363)(23,1.850)(24,1.507)(25,1.363)
(26,1.130)(27,1.500)(28,1.530)(29,0.643)(30,1.108)
(31,1.096)(32,1.821)(33,1.641)(34,1.099)(35,1.091)
(36,1.091)(37,1.967)(38,1.604)(39,1.414)(40,1.382)
(41,1.420)(42,1.092)(43,1.060)(44,1.515)(45,1.148)
(46,1.110)(47,1.105)(48,2.094)(49,1.592)(50,1.125)
(51,0.813)(52,1.121)(53,1.919)(54,1.701)(55,1.921)
(56,1.689)(57,1.145)(58,0.988)(59,1.320)(60,2.500)
(61,1.715)(62,0.956)(63,1.366)(64,1.498)(65,1.406)
(66,1.349)(67,1.382)(68,1.756)(69,3.123)(70,2.908)
(71,2.081)
};
\addplot coordinates{};
\end{axis}
\end{tikzpicture}
\end{minipage}

    \caption{Speedup ratio of all 71 benchmarks compared between \mysys and SABRE in four architectures. The benchmarks are listed from left to right in the ascending order of the number of qubits used.}
    \label{fig:result}
\end{figure}
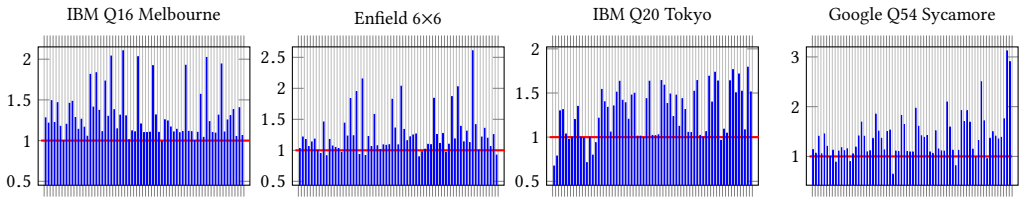
\mysect{Circuit Execution Speedup}
We collect the weighted circuit depth of the circuits produced by \mysys and SABRE for the 71 benchmarks. Initial mapping has been proved to be significant for the qubit mapping problem, and for a fair comparison, we use the same method as SABRE to create the initial mapping for the benchmarks. 
We use the depth of circuits produced by SABRE compared with the one of \mysys to show the ability of our algorithm to speed up the quantum program. 
As shown in Fig.\ref{fig:result}, 
the average speedup ratio of \mysys on four architecture models, IBM Q16 Melbourne, Enfield 6$\times$6, IBM Q20 Tokyo and Google Q54 are respectively 1.212, 1.241, 1.214 and 1.258.

\section{Conclusion}
\label{sec:concl}
In NISQ era, most quantum programs are not directly executable 
because two-qubit gates can be applied between arbitrary two logical qubits 
while it can only be implemented between two adjacent physical qubits due to hardware constraints. 
To solve this problem,
in this paper we propose \mysys that can transform the origin circuit and insert necessary \kn{SWAP} operations 
making the circuit comply with the hardware constraints. 
With the design of qubit lock and commutativity detection, 
\mysys is aware of program context and the gate duration difference 
which help \name find the remapping with good parallelism and
reduce QC's weighted depth. 
Experimental results show that 
compared to the best known remapping algorithm, 
\name can cut down 17.5\% $\sim$ 19.4\% weighted depth at average.


\iftrue

\else
\bibliography{quantum}
\bibliographystyle{plain}
\fi
\end{document}